\pdfoutput=1
\documentclass[12pt]{article}

\usepackage[parfill]{parskip}
\usepackage[colorlinks=true]{hyperref}
\usepackage[margin=1in]{geometry}
\usepackage{authblk}
\usepackage{amsmath}
\usepackage{verbatim}
\usepackage{graphicx}
\usepackage{xcolor}
\usepackage{listings}
\usepackage{courier}
\usepackage{fancyvrb}

\definecolor{mygreen}{rgb}{0,0.6,0}
\definecolor{mygray}{rgb}{0.5,0.5,0.5}
\definecolor{mymauve}{rgb}{0.58,0,0.82}
\definecolor{myred}{rgb}{0.7,0,0}
\definecolor{myblue}{rgb}{0,0,0.7}

\title{\bf OpenSesame tutorial}
\author[1]{Daniel A. Rehn}
\author[2]{Carl W. Greeff}
\author[1]{Daniel G. Sheppard}
\author[1]{Charles E. Starrett}
\author[2]{Scott D. Crockett}
\affil[1]{\small Computational Physics Division, Los Alamos National Laboratory}
\affil[2]{\small Theoretical Division, Los Alamos National Laboratory}
\date{April, 2021}

\begin{document}
\maketitle

\begin{abstract}
  OpenSesame is a program for generating tabular equations of state
  (EOS), with capabilities for multiphase EOS construction. In this
  tutorial, we provide an overview of how to run OpenSesame to
  construct a multiphase EOS. We discuss some general features of
  OpenSesame, followed by a description of sample input files required
  for multiphase EOS construction. We also discuss how to extract data
  from EOS tables in order to compare to experimental data, with an
  example using the OpenSesame GUI. Lastly, we provide a description
  of how to generate ASCII-formatted EOS tables most often used by
  hydro code users.
\end{abstract}
\newpage

\section*{Overview}
OpenSesame~\cite{george0} is a program for generating tabular
equations of state (EOS) with support for the construction of
multiphase EOS~\cite{chisolm2005constructing}. Multiphase EOS are the
main focus of this tutorial.

The basic thermodynamic variable used in OpenSesame is the Helmholtz
free energy, denoted $F$ and sometimes $A$.  OpenSesame relies on a
decomposition of the total free energy into 3 pieces:

\begin{equation}
  F(V,T) = E_c(V) + F_{\rm ion}(V,T) + F_{\rm el}(V,T).
  \label{eq:F}
\end{equation}

These terms are, from left to right, the \emph{cold curve} or static
lattice energy, the free energy associated with ionic motion, and the
electronic free energy. Each term is computed using a different
materials model, and multiple choices of materials models are
available for each term. The choice of materials models are up to the
user and depend upon the material phase (solid, liquid, gas, etc.), as
well as the user's particular preferences in their choice of models.
In each case, the models are evaluated over a range of $V$ and $T$ and
stored in tables (tabular format).  Those tables can then be
interpolated to provide data, such as isotherms, Hugoniots, etc., that
can be compared to experiments.

\begin{figure}[!ht]\centering
  \includegraphics[width=0.7\textwidth]{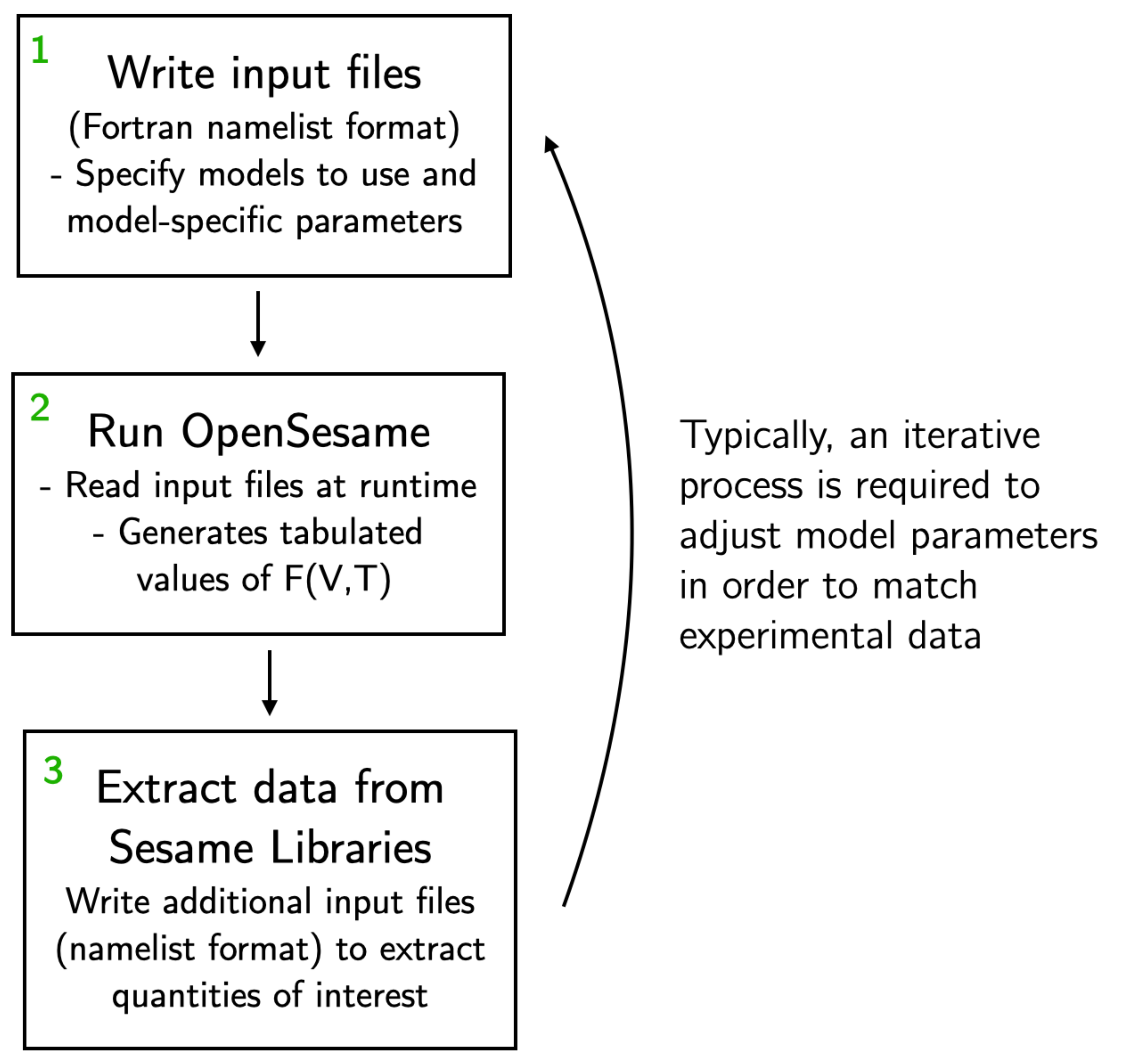}
  \caption{\footnotesize Three main steps in running OpenSesame}
  \label{fig:steps}
\end{figure}

Once $F(V,T)$ is computed, other thermodynamic quantities can be
evaluated using $F(V,T)$. Some examples include the
following~\cite{crockett},

\begin{align*}
  E &= F + TS = F - T\left({\partial F\over \partial T}\right)_V  &  \text{internal energy}\\
  S &= -\left({\partial F \over \partial T}\right)_V & \text{entropy}\\
  P &= \rho^2 \left({\partial F \over \partial \rho}\right)_T & \text{pressure}\\
  C_V &= -T\left({\partial^2 F\over \partial T^2}\right)_V =
         \left({\partial E \over \partial T}\right)_V =
         T\left({\partial S \over \partial T}\right)_V
         &\text{specific heat at constant volume}\\
  C_P &= \left({\partial H \over \partial T}\right)_P =
        T\left({\partial S \over \partial T}\right)_P
        &\text{specific heat at constant pressure}\\
  B_T &= V\left({\partial^2 F \over \partial V^2}\right)_T =
         -V\left({\partial P \over \partial V}\right)_T &
         \text{isothermal bulk modulus}\\
  \alpha &= -{1\over V}\left({\partial V \over \partial T}\right)_P &
  \text{thermal expansion coefficient}\\
  \Gamma &= V\left({\partial P \over \partial E}\right)_V &
  \text{Gr\"{u}neisen parameter}
\end{align*}
\newpage
Running OpenSesame involves three main steps, shown schematically in
Fig.~\ref{fig:steps}. Typically, these steps are repeated in a
cyclical manner so that model parameters can be adjusted to provide
good agreement with experimental data.  In the following sections, we
discuss each of these 3 steps in detail.

Unless otherwise noted, units used in OpenSesame are those in
Table~\ref{tab:units}.

\begin{table}[!ht]\centering
  \begin{tabular}{|l|l|l|}\hline
    Variable      & Description                         & Units   \\ \hline
    $\rho$ (R)    & density                             & g/cm$^3$\\
    $V$           & specific volume                     & cm$^3$/g\\
    $T$           & temperature                         & K       \\
    $P$           & pressure                            & GPa     \\
    $E$, $F$, $A$, $G$ & energy (Internal, Helmholtz, Gibbs) & MJ/kg   \\
    $C_V$, $C_P$  & specific heat                       & MJ/kg/K \\
    $S$           & entropy                             & MJ/kg/K \\
    $B_T$, $B_S$  & bulk modulus                        & GPa     \\
    $u$, $c_s$    & velocity, sound speed               & km/s    \\
    $\alpha_v$  (alpha) & volume thermal expansion coefficient & 1/K \\
    $\mu$ (mu), G (muM) & shear modulus at T=0, shear modulus at melt & GPa\\ \hline
  \end{tabular}
  \caption{\footnotesize Variable names and units used in OpenSesame}
  \label{tab:units}
\end{table}

\begin{figure}[!ht]\centering
  \includegraphics[width=0.8\textwidth]{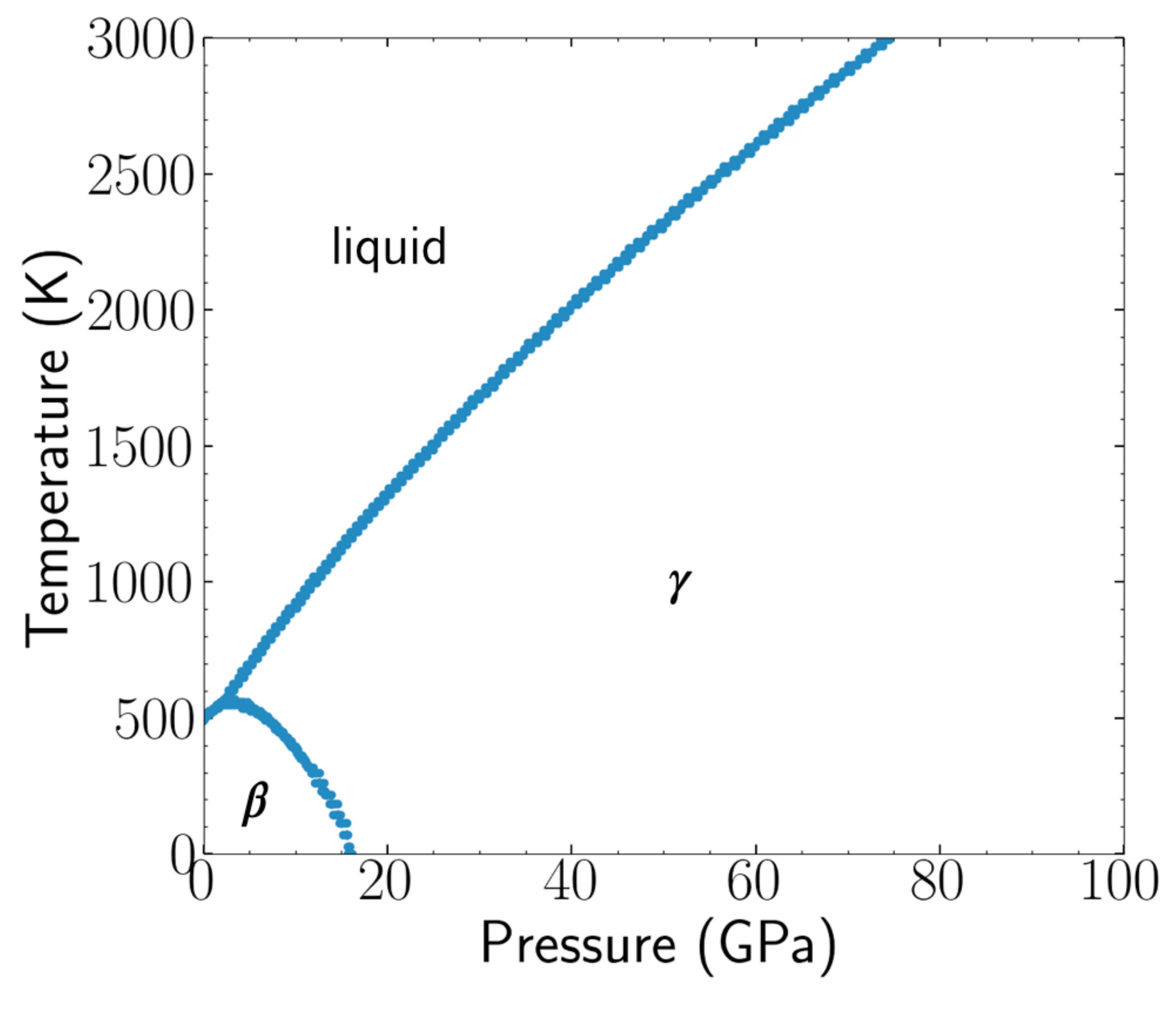}
  \caption{\footnotesize Phase diagram of tin from SESAME 2161. $\beta$ and $\gamma$
    are the solid phases}
  \label{fig:pt}
\end{figure}

Below, we will use SESAME 2161 (material = tin) as an example.  SESAME
2161 was developed by Carl Greeff~\cite{greeff2005sesame} and is a
multiphase EOS with 2 solid phases (labeled $\beta$ and $\gamma$) and
the liquid phase. The phase diagram of SESAME 2161 is shown in
Fig.~\ref{fig:pt}.

Note that a new tin EOS, SESAME 2162, has also been developed and
includes 4 solid phases of
tin~\cite{rehn2021multiphase,rehn2020using}, and therefore can be
thought of as the successor to 2161. However, for tutorial purposes,
it is easier to discuss the simpler case of just 2 solid phases.

\newpage

\section*{1. Writing OpenSesame input files}
OpenSesame relies on input files in Fortran namelist
format~\cite{metcalf2011modern}. Although it is possible to write just
one input file for a full multiphase EOS, it is sometimes preferable
to split the input files into parts that correspond to the individual
phases.  Here, we show a typical directory structure for a multiphase
EOS,

\begin{Verbatim}[fontsize=\footnotesize]
$ ls
1-beta/    2-gamma/    3-liquid/    4-multiphase/    lib/    plots/    run.sh
\end{Verbatim}

Folders \texttt{1-3} contain input files for the individual phase
tables, while \texttt{4-multiphase} contains the input file for the
multiphase construction.  The \texttt{lib} folder contains the tables
produced by running the input files in folders \texttt{1-4} and
\texttt{plots} includes plotting scripts, discussed later. Lastly,
\texttt{run.sh} is a bash script that runs OpenSesame in each folder.

The contents of these folders is as follows,

\begin{Verbatim}[fontsize=\footnotesize]
$ ls *
1-beta:
input.nml

2-gamma:
input.nml

3-liquid:
input.nml

4-multiphase:
input.nml

lib:
002101/    002102/    002103/    002104/    002161/    material_directory
\end{Verbatim}

The folders \texttt{1-4} contain only an \texttt{input.nml}
file. After running \texttt{opensesame}, other ASCII format output
files are generated in these directories. The \texttt{lib} directory
contains folders that store tabulated thermodynamic quantities.  The
numbers (\texttt{002101}, \texttt{002102}, etc.) of these directories
correspond to each phase, and the numbers themselves are specified in
the \texttt{input.nml} files. More details on \texttt{lib} are provided
later in the tutorial.

\newpage

\subsection*{$\beta$ phase input}
The input file for the $\beta$ phase in \texttt{1-beta} is copied in
Listing~\ref{lst:beta}. Note the comments in green font.  The cold
curve model is the Rose-Vinet model~\cite{vinet1987temperature}, the
ionic model is the Debye model~\cite{debye}, and the electronic model
is the Thomas-Fermi-Dirac
model~\cite{chisolm2003thomas,fermi1927metodo,thomas1927calculation}.

\begin{lstlisting}[language=Fortran,caption={\footnotesize\texttt{1-beta/input.nml}:
  input file for $\beta$ phase.}, label={lst:beta}]
&job 
    job_type = 'neweos'         ! must select job type, create new EOS
    resultlib_path = '../lib'
/
&neweos                         ! define parameters for new EOS
    material_number      = 2101 ! specify mat # (corresponds to lib/002101) 

    ! basic material properties
    material_name        = 'tin'
    author               = 'C. Greeff & E. Chisolm, T-1'
    atomic_number        = 50
    atomic_weight        = 118.710
    reference_density    = 7.28729
    ifstandard           = .false.  ! nuanced options, see manual
    match_low            = .false.

    ! cold curve parameters
    cold_model           = 'rose'
    cold_density         = 7.42402
    cold_bulk_modulus    = 57.0
    cold_dbdp            = 4.9

    ! ion model parameters
    nuclear_model        = 'debye'
    gruneisen_option     = 7
    reference_debye      = 144.306
    reference_gamma      = 2.2
    dgamma_left          = -2.2
    dgamma_right         = -2.2

    ! electronic model parameters
    electron_model       = 'tfd' 
    elec_low_temp_interp = .true.
/
\end{lstlisting}

\newpage
\subsection*{$\gamma$ phase input}
The input file for the $\gamma$ phase in \texttt{2-gamma} is shown in
Listing~\ref{lst:gamma}. Again, note the descriptive comments in green
font. The models used for $\gamma$ can be the same as those of the
$\beta$ phase, but here \texttt{hightliq} is used for some complicated
reasons related to the re-emergence of $\gamma$ phase in the liquid region.

\begin{lstlisting}[language=Fortran,caption={\footnotesize\texttt{2-gamma/input.nml}:
  input file for the $\gamma$ phase.}, label={lst:gamma}]
&job 
    job_type = 'neweos' 
    resultlib_path = '../lib'
/

&neweos 
    material_number          = 2102  ! corresponds to lib/002102

    ! basic material properties
    material_name            = 'tin' 
    author                   = 'C. Greeff & E. Chisolm, T-1'
    atomic_number            = 50 
    atomic_weight            = 118.710
    reference_density        = 7.02426
    ifstandard               = .false.
    match_low                = .false.

    ! cold curve parameters
    cold_model               = 'rose'
    cold_density             = 7.51329
    cold_bulk_modulus        = 50.0
    cold_dbdp                = 5.6
    energy_shift             = 0.0387499

    ! ion model parameters
    nuclear_model            = 'hightliq'  ! or 'debye'
    gruneisen_option         = 7
    reference_debye          = 90.7148
    reference_gamma          = 3.2
    dgamma_left              = -4.48
    dgamma_right             = -4.48

    ! electronic model parameters
    electron_model           = 'tfd' 
    elec_low_temp_interp     = .true.

    ! melt parameters if using hightliq
    melt_option              = 'lindemann'
    initial_melt_density     = 7.595
    initial_melt_temperature = 650.0
    liquid_shift             = .true.
    entropy_difference       = 0.0
/
\end{lstlisting}

\newpage
\subsection*{Liquid phase input}
\vspace{-0.1in}
The input file for the liquid phase in \texttt{3-liquid} is shown in
Listing~\ref{lst:liquid}. Again, note the descriptive comments in
green. The ionic model used here is described in
Refs.~\cite{chisolm2003test,chisolm2005extending}.

\begin{lstlisting}[language=Fortran,caption={\footnotesize\texttt{3-liquid/input.nml}:
    input file for the liquid phase.},label={lst:liquid}]
&job job_type='neweos' 
    resultlib_path='../lib'
/

&neweos
    material_number          = 2103  ! corresponds to lib/002103

    ! basic material properties
    material_name            = 'tin' 
    author                   = 'C. Greeff & E. Chisolm, T-1'
    atomic_number            = 50 
    atomic_weight            = 118.710
    ifstandard               = .false.

    ! cold curve parameters
    cold_model               = 'rose'
    cold_density             = 7.51329
    cold_bulk_modulus        = 50.0
    cold_dbdp                = 5.6
    energy_shift             = 0.0387499

    ! ion model parameters
    nuclear_model            = 'hightliq'
    gruneisen_option         = 7
    reference_debye          = 90.7148
    reference_gamma          = 3.2
    dgamma_left              = -4.48
    dgamma_right             = -4.48

    ! electronic model parameters
    electron_model           = 'tfd' 
    elec_low_temp_interp     = .true.

    ! melt parameters
    melt_option              = 'lindemann'
    initial_melt_density     = 7.595
    initial_melt_temperature = 650.0
    liquid_shift             = .true.
    entropy_difference       = 0.5412

    ! additional parameters related to behavior in expansion
    reference_density        = 7.02426
    cohesive_energy          = 2.3904
    lennard_jones_exponent   = 0.9
/
\end{lstlisting}
\newpage

\subsection*{Multiphase input}
The multiphase file is long, so we break it into 2 main parts:
\begin{enumerate}
\item Grid construction
\item Multiphase construction
\end{enumerate}

\subsubsection*{Grid construction}
The grid consists of two parts: a compression ($\rho/\rho_0$) grid and
temperature ($T$) grid.  The compression grid is defined in
Listing~\ref{lst:comp-grid}.

\begin{lstlisting}[language=Fortran,caption={\footnotesize\texttt{4-multiphase/input.nml}:
    defining the compression grid}, label={lst:comp-grid}]
&job job_type = 'grid' /

&grid 
    grid_type   = 'compression' 
    grid_change = 'new'  ! create a new compression grid
    grid_size   = 85
    grid_new    = 0.0000e+00, 0.1000e-05, 0.2000e-05, 0.5000e-05, 0.1000e-04, 
                  ...
                  0.8000e+02, 0.9000e+02, 0.1000e+03, 0.1250e+03, 0.1500e+03
/

&job job_type = 'grid' /

&grid
    grid_type     = 'compression'
    grid_change   = 'insert'   ! insert a grid on the existing compression grid
    !grid_file     = ''  ! option to import a grid from a file
    grid_spec     = 'lin'
    grid_size     = 420
    grid_limit_lo = 0.9600e+00
    grid_limit_hi = 0.2100e+01
/
\end{lstlisting}

Note that the first grid is specified with
\texttt{grid\_change='new'}, generating a new compression grid and the
values are defined manually with different compression values (some
lines are left out for brevity, denoted by \texttt{...}).  After this,
the second grid specification is of type \texttt{'insert'}, which adds
a linearly spaced grid of 420 points between 0.96 and 2.1.  This
option is mostly used to provide a refined grid of points over
particularly relevant compression ranges.

\newpage
After defining the compression grid, we define the temperature grid in
a similar way, shown in Listing~\ref{lst:temp-grid}.

\begin{lstlisting}[language=Fortran,caption={\footnotesize\texttt{4-multiphase/input.nml}:
    defining the temperature grid}, label={lst:temp-grid}]
&job job_type              = 'grid' /
&grid 
    grid_type              = 'temperature' 
    grid_change            = 'new'  ! construct a new grid
    grid_temperature_units = 'ev'
    grid_file              = ''
    grid_size              = 54
    grid_new               =  
      0.0000e+00, 0.2500e-02, 0.6250e-02, 0.1000e-01, 0.1250e-01, 
      ...
      0.8000e+03, 0.9000e+03, 0.1000e+04, 0.1500e+04, 0.2500e+04, 
/

&job job_type              = 'grid' /

&grid
    grid_type              = 'temperature'
    grid_change            = 'insert'   ! add to existing grid
    grid_temperature_units = 'ev'
    grid_size              = 25
    grid_spec              = 'lin'
    grid_limit_lo          = 0.2750e-01
    grid_limit_hi          = 0.5000e-01 
/

!  some additional 'insert' grids are omitted for brevity here
\end{lstlisting}
Note here that we again use a \texttt{'new'} grid and an
\texttt{'insert'} grid to refine it, with additional \texttt{'insert'}
grids left out for brevity. Temperatures can be specified in eV or K.

\newpage
\subsubsection*{Multiphase construction}
After defining the grids, we turn to the multiphase construction.  The
multiphase construction is broken into two parts: an initial
"non-standardized" table called 2104, followed by a "standardized"
table called 2161. The standardization step is described below.

The primary multiphase construction input is shown in Listing~\ref{lst:mp}.

\begin{lstlisting}[language=Fortran,caption={\footnotesize\texttt{4-multiphase/input.nml}:
     construction of the non-standardized 2104 table},label={lst:mp}]
&job
  job_type = 'materials' 
  sourcelib_path = '../lib' 
  resultlib_path = '../lib'
/

&materials
    author              = 'C. Greeff & E. Chisolm, T-1'
    references          = 'LA-UR-05-9414'
    material_option     = 'multiphase'
    result_material     = 2104
    reference_density   = 7.28729

    ! specify material ids for 2104 creation
    nmats               = 3
    source_materials(1) = 2101
    source_materials(2) = 2102
    source_materials(3) = 2103
    phase_names(1)      = 'beta'
    phase_names(2)      = 'gamma'
    phase_names(3)      = 'liquid'

    ! windows to limit range of given phases
    ph_rholo(1)         = 6.0
    ph_rhohi(1)         = 10.0
    ph_thi(1)           = 1000.0 
    ph_rholo(2)         = 7.0
/
\end{lstlisting}

The 2104 multiphase table is built from the 3 phase tables described
above, with numbers \texttt{2101,2102,2013}. Also note the last four
lines: these variables specify "windows" that determine the min and
max values of density and temperature for each phase. The point of the
windows is to prevent the reappearance of certain phases in regions of
the phase diagram where they don't belong.  For example, here
\texttt{ph\_rholo(2)} says that the minimum density of the gamma phase
(material index 2 defined above) is 7 g/cm$^3$. Therefore, in the
multiphase construction, gamma phase cannot reappear at densities
below 7 g/cm$^3$.

\newpage
After the definition of the multiphase 2104 table, we create the 2161
EOS via a ``standardization'' step, described above.  The
standardization step simply resets the zero of energy to a value at
room $T$ (298.15 K) and ambient pressure.  This is done because hydro
codes prefer to define the 0 of energy to ambient conditions. The
input lines to do this are as in Listing~\ref{lst:standardize}.

\begin{lstlisting}[language=Fortran,caption={\footnotesize\texttt{4-multiphase/input.nml}:
    standardization step to create the 2161 table.}, label={lst:standardize}]
!  Standardization step
&job job_type           = 'materials'
    sourcelib_path      = '../lib'
    resultlib_path      = '../lib'
/

&materials
    author              = 'Eric Chisolm, T-1'
    material_option     = 'standardize'
    source_materials    = 2104 
    result_material     = 2161
/
\end{lstlisting}

After standardization the 2161 table itself could be considered
"complete''. However, depending on the needs of the users, other
options for additional tables to add to the 2161 EOS are
available. These include,

\begin{enumerate}
\item Maxwell construction
\item Construction of melt curve tables
\item Generation of shear modulus tables
\end{enumerate}

These do not necessarily have to be done.  Here we focus only on melt
curve table generation, which is frequently desired by hydro code
users.

\newpage
The input for melt curve generation is shown in
Listing~\ref{lst:melt}.

\begin{lstlisting}[language=Fortran,caption={\footnotesize\texttt{4-multiphase/input.nml}:
      generation of melt curve tables},label={lst:melt}]
!  Generating melt curves    
&job
    job_type       = 'neweos' 
    sourcelib_path = '../lib'
    resultlib_path = '../lib'
/

&neweos
    material_number               = 2161
    material_name                 = 'tin'
    author                        = 'C. Greeff & E. Chisolm, T-1'
    atomic_number                 = 50
    atomic_weight                 = 118.710
    reference_density             = 7.28729
    melt_gamma_reference_density  = 7.02426
    melt_model                    = 'multiphase'
    melt_option                   = 'lindemann'
    melt_gruneisen_option         = 7 
    melt_reference_debye          = 90.7148
    melt_reference_gamma          = 3.2
    melt_dgamma_left              = -4.48
    melt_dgamma_right             = -4.48 
    initial_melt_density          = 7.595
    initial_melt_temperature      = 650.0
    solid_multiplier              = 0.99
    liquid_multiplier             = 1.00 
/
\end{lstlisting}

Parameters for the melt table are basically the same as the liquid
phase, but notice we specify some \texttt{melt\_} variable options.
These behave in the same way as the liquid parameters, it is just that
here the point is to generate melt curves in a different format of
$(T,P)$ points or $(\rho,T)$ points.

\newpage
\section*{2. Running OpenSesame}
I recommend installing OpenSesame on a personal laptop because it runs
much faster than on the HPC machines and the data storage is then
local. After building the source code, you can add the location of the
binaries to your \texttt{PATH} variable in a \texttt{~/.bashrc} (or
analogous) file.  At that point, the following binaries should be
available:

\begin{Verbatim}[fontsize=\footnotesize]
/path/to/OpenSesame/OpenSesameSource/opensesame
/path/to/OpenSesame/OpenSesameSource/gui/opensesamegui.real
\end{Verbatim}

The first of these can be run in each directory where an
\texttt{input.nml} file is located, for example via,

\begin{Verbatim}[fontsize=\footnotesize]
$ opensesame < input.nml
\end{Verbatim}

The GUI executable requires slightly more work to set up. For this,
you will need to create an additional executable file located at

\begin{Verbatim}[fontsize=\footnotesize]
/path/to/OpenSesame/OpenSesameSource/gui/opensesamegui
\end{Verbatim}

The contents of this file consist of just one line:

\begin{Verbatim}[fontsize=\footnotesize]
echo "source /path/to/OpenSesame/OpenSesameSource/gui/opensesamegui.real ; main" | wish
\end{Verbatim}

Then you will want to allow executable permissions on that file:

\begin{Verbatim}[fontsize=\footnotesize]
$ chmod u+x /path/to/OpenSesame/OpenSesameSource/gui/opensesamegui
\end{Verbatim}

It is then possible to run \texttt{opensesamegui} from the command
line. We will show examples of using the GUI to plot different
quantities in Part 3 below.

\newpage
\section*{\texttt{run.sh} script}
\vspace{-0.1in} Recall the directory structure from above, where we see a
\texttt{run.sh} script,
\begin{Verbatim}[fontsize=\footnotesize]
$ ls 
 1-beta/   2-gamma/   3-liquid/   4-multiphase/   lib/   plots/   run.sh
\end{Verbatim}

The contents of \texttt{run.sh} are shown in Listing~\ref{lst:run}.
\begin{lstlisting}[language=Bash,caption={\footnotesize\texttt{run.sh} script},label={lst:run}]
#!/bin/bash
SESAME=opensesame
# create lib folder for output
[ ! -d './lib' ] && mkdir lib
{
  folder='1-beta'
  echo '-> running in' $folder;
  ti=`date +%s`;
  cd $folder;
  $SESAME < input.nml > output
  tf=`date +%s`;
  echo '-> ' $folder ' runtime (seconds) = ' $((tf-ti))
} &
wait;  # optionally take this out to run 1-beta, 2-gamma simultaneously
{
  folder='2-gamma'
  echo '-> running in' $folder;
  ti=`date +%s`;
  cd $folder;
  $SESAME < input.nml > output
  tf=`date +%s`;
  echo '-> ' $folder ' runtime (seconds) = ' $((tf-ti))
} &
wait;  # optionally take this out to run 2-gamma, 3-liquid simultaneously
{
  folder='3-liquid'
  echo '-> running in' $folder;
  ti=`date +%s`;
  cd $folder;
  $SESAME < input.nml > output
  tf=`date +%s`;
  echo '-> ' $folder ' runtime (seconds) = ' $((tf-ti))
} &
wait;  # do not remove: multiphase must be done after all 3 phases finish
{
  folder='4-multiphase'
  echo '-> running in' $folder;
  ti=`date +%s`;
  cd $folder;
  $SESAME < input.nml > output
  tf=`date +%s`;
  echo '-> ' $folder ' runtime (seconds) = ' $((tf-ti))
} &
wait;
\end{lstlisting}

Note that the \texttt{wait} statements after the $\beta$ and $\gamma$
cases can in principle be removed to allow the $\beta$, $\gamma$ and
liquid phase construction to occur in parallel. However, the
multiphase construction must be done after the completion of the
individual phases, since it relies on those tables already being
present to form the 2104 table.

\newpage
\section*{3. Extract data from Sesame libraries}
We now describe how to extract data from the OpenSesame output so that
comparisons to relevant experiments can be made. There are two main
parts to this section,
\begin{enumerate}
\item A description of the types of data contained in the \texttt{lib}
  folder
\item How to run opensesame to extract data from \texttt{lib} and
  compare to experimental data
\end{enumerate}

Step 2 can be done by hand, using \texttt{opensesame} directly, or it
can be done using the \texttt{opensesamegui}. We will look at the
\texttt{opensesamegui} way first, since this method automatically
generates opensesame input files that can be then be used to plot
data.

\subsection*{Layout of the \texttt{lib} directory}
As shown earlier, the \texttt{lib} folder consists of the following,

\begin{Verbatim}[fontsize=\footnotesize]
$ ls lib
002101/   002102/   002103/   002104/   002161/   material_directory
\end{Verbatim}

Each of the numbered directories contains subdirectories. The
\texttt{2161} table is of most interest, so we will look at its
subdirectories. Note that other individual phase tables are laid out
similarly. The subdirectories of \texttt{2161} are,

\begin{Verbatim}[fontsize=\footnotesize]
$ ls lib/002161
101     103     311/     401/     412/     432/
102     301/    321/     411/     431/     table_directory
\end{Verbatim}

Here we see both files and folders labeled by different numbers. The
numbers have a specific meaning that are listed in
\texttt{table\_directory}, with a brief description
here~\cite{crockett,george}:

\begin{itemize}
\item 100 series (comments)
  \begin{itemize}
  \item 101: provides basic information: name, authors, etc.
  \item 102-199: free-form text describing an other information
  \end{itemize}
\item 201: atomic number, atomic weight, reference density, etc.
\item 300 series (thermodynamic functions on a $\rho, T$ grid)
  \begin{itemize}
  \item 301: total $\rho, T, P, E, F$
  \item 303: cold + nuclear only
  \item 304: electronic only
  \item 305: ionic only
  \item 306: cold curve $\rho, T, P, E, F$
  \item 311: Maxwell constructed 301 table
  \item 321: mass fractions for multiphase EOS (used to plot phase
    diagram)
  \end{itemize}
\item 400 series (functions along a curve)
  \begin{itemize}
  \item 401: vapor dome $\rho, T, P, E, F$
  \item 411: solidus $\rho, T, P, E, F$
  \item 412: liquidus $\rho, T, P, E, F$
  \item 431: shear modulus at $T=0$
  \item 432: shear modulus at $T=0$ and $T=T_{\rm melt}$
  \end{itemize}
\item 500 series (opacities)
\item 600 series (conductivity)
\end{itemize}

If we now look inside, for example the \texttt{301} (total) folder, we
see the following:

\begin{Verbatim}[fontsize=\footnotesize]
$  ls 002161/301/
A      E      P       R       T       item_directory
\end{Verbatim}

In short, each of these are files that contain different thermodynamic
quantities on the defined grids. Basic information about these files,
e.g., their sizes and number of points, etc. are found in the
\texttt{item\_directory} file.

\subsection*{Extract data from \texttt{lib} to compare with experiments}
The tables themselves should not be used directly. Instead, we want to
generate additional input files that run OpenSesame to extract useful
quantities from these tables. In this way, OpenSesame does all of the
interpolation and data handling for us. The easiest way to extract
data is by running the OpenSesame GUI. Running the GUI automatically
creates OpenSesame input files that can subsequently be used to
generate data for plotting.

To demonstrate use of the GUI, we return to the root directory of EOS
we are working on and create a new folder called \texttt{gui}, then we
move to that directory and execute \texttt{opensesamegui},

\begin{Verbatim}[fontsize=\footnotesize]
$ ls
1-beta/    2-gamma/    3-liquid/    4-multiphase/    lib/    plots/    run.sh
$ mkdir gui
$ cd gui
$ opensesamegui
\end{Verbatim}

After running this, a GUI window should pop up. It may require X
Windows (on Mac, XQuartz) to be installed. The following pages will be
completely graphical, showing how to use the GUI.

\newpage
\subsection*{Running the GUI}
\begin{center}  \includegraphics[width=\textwidth]{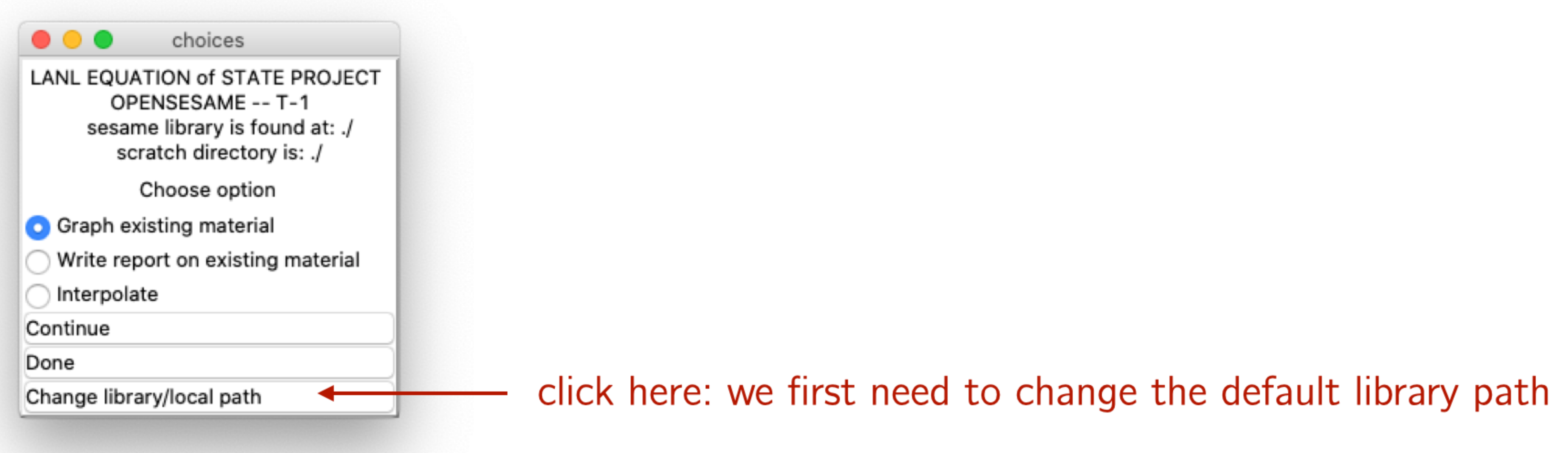} \end{center}
\begin{center}  \includegraphics[width=\textwidth]{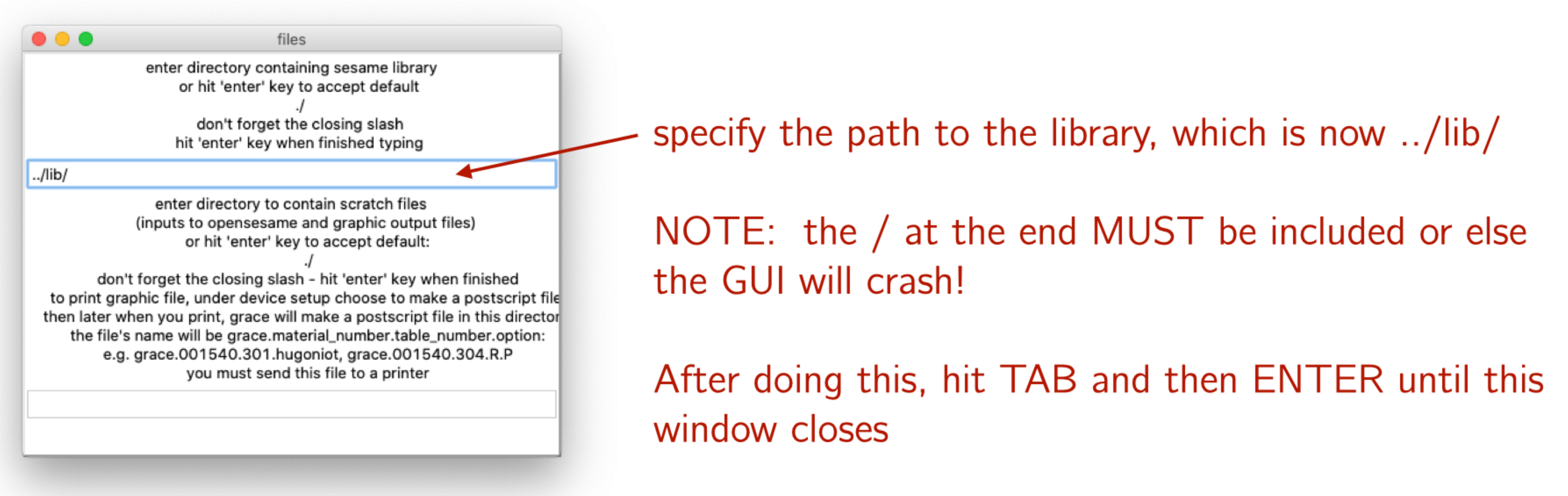} \end{center}
\begin{center}  \includegraphics[width=\textwidth]{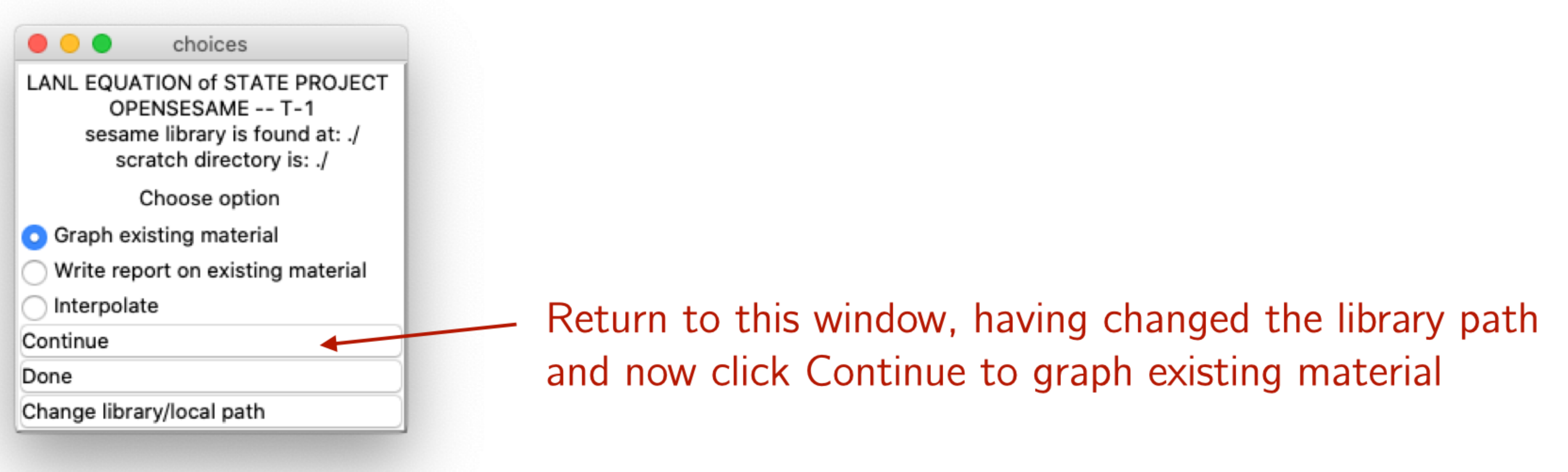} \end{center}
\begin{center}  \includegraphics[width=\textwidth]{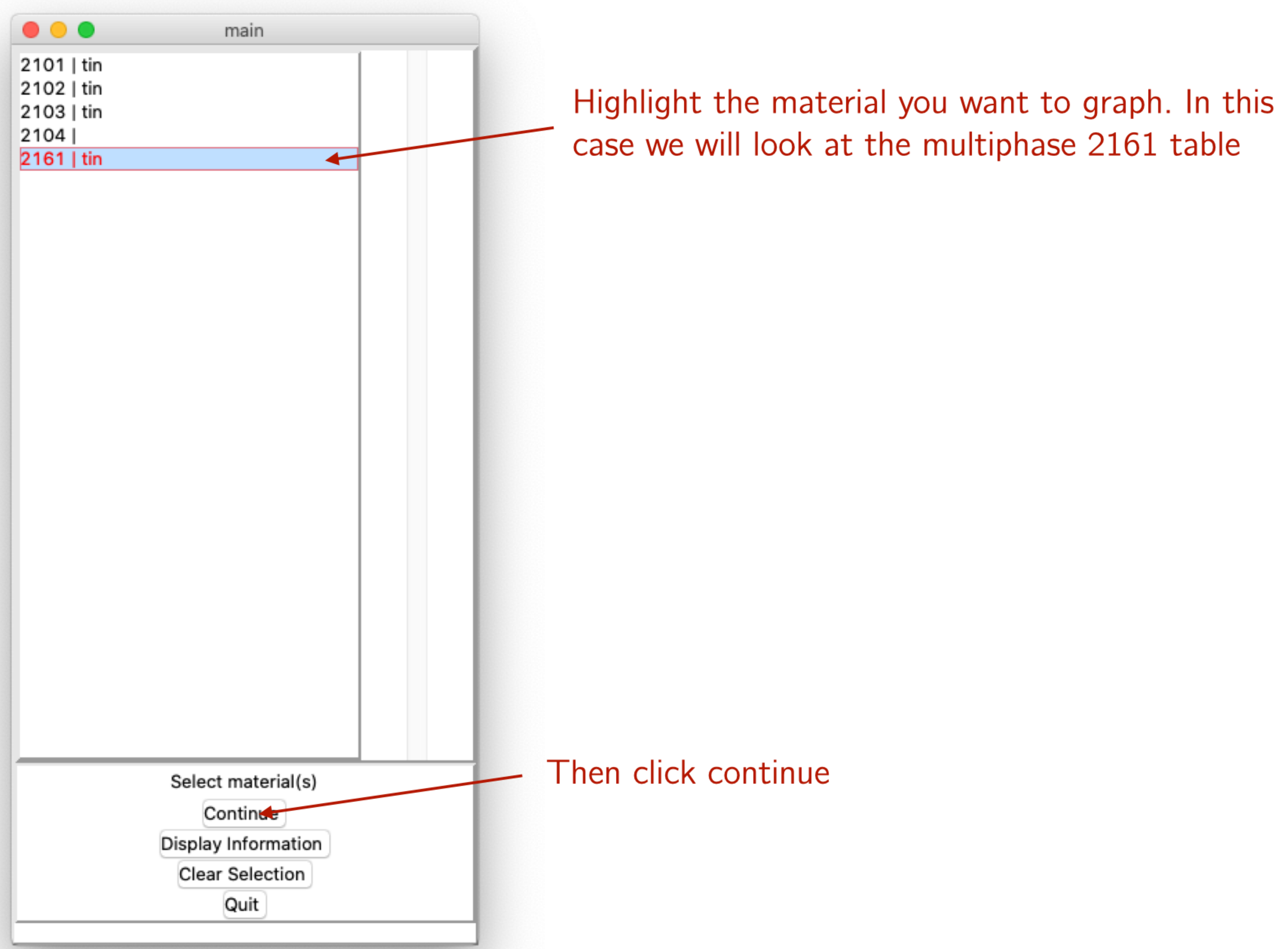} \end{center}
\begin{center}  \includegraphics[width=\textwidth]{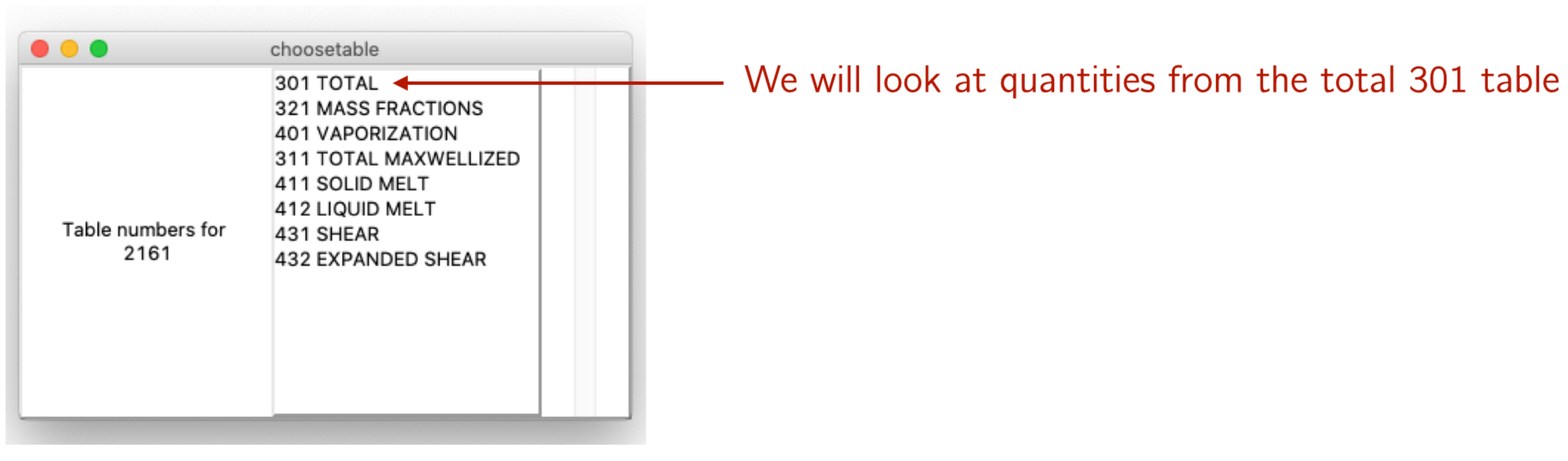} \end{center}
\begin{center}  \includegraphics[width=\textwidth]{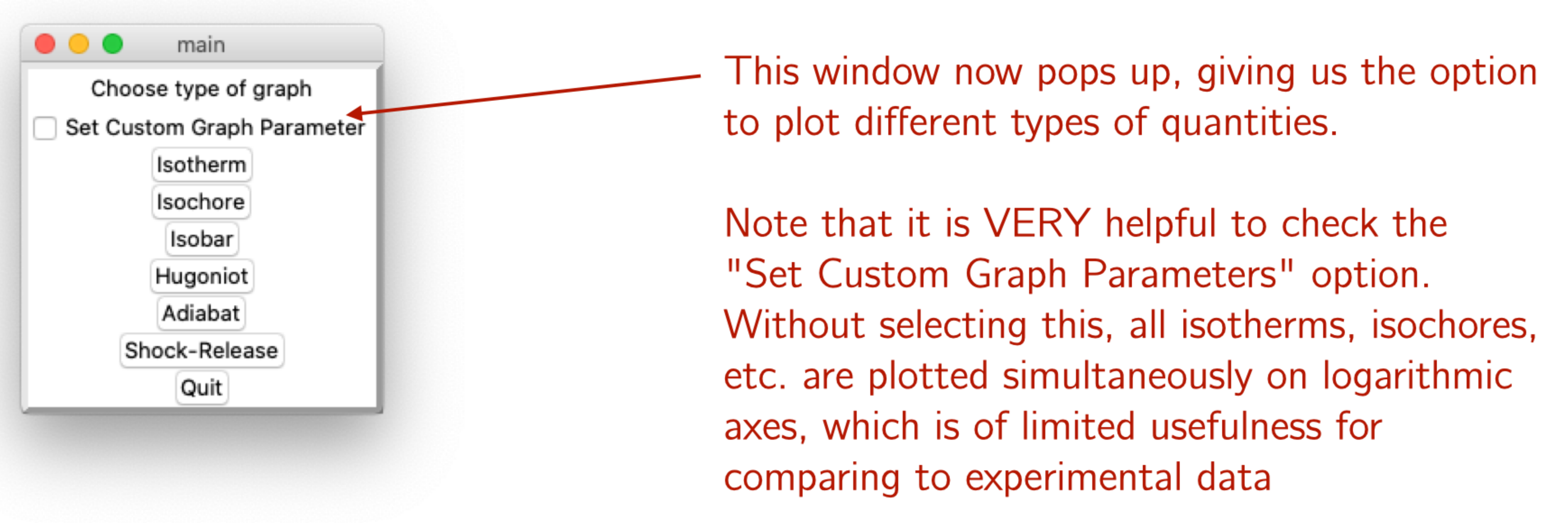} \end{center}
\begin{center}  \includegraphics[width=\textwidth]{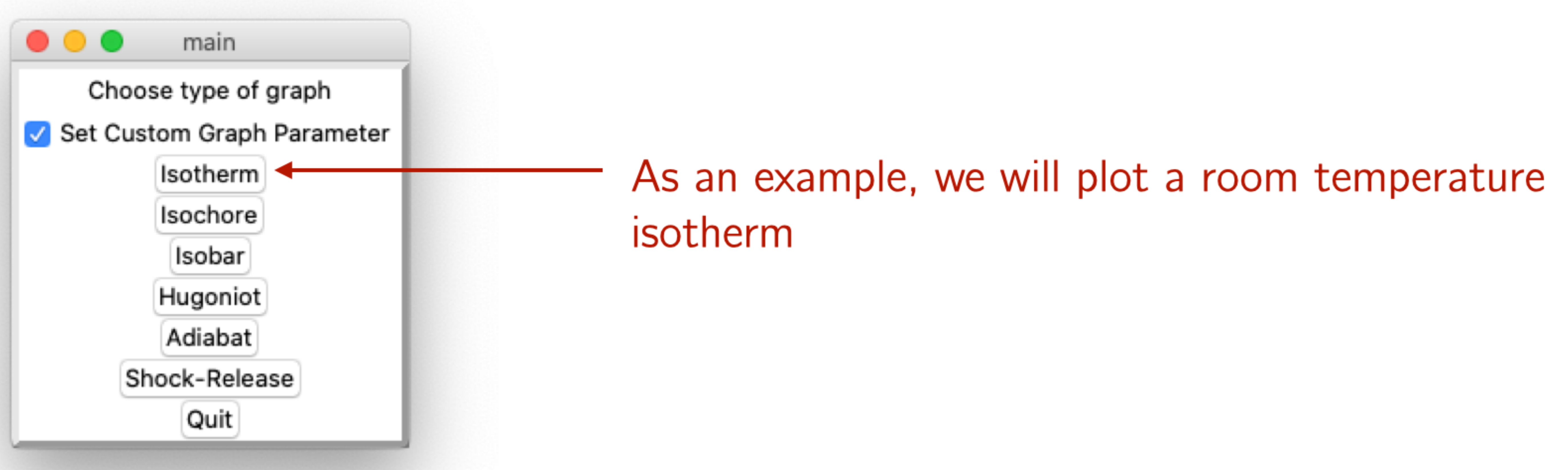} \end{center}
\begin{center}  \includegraphics[width=\textwidth]{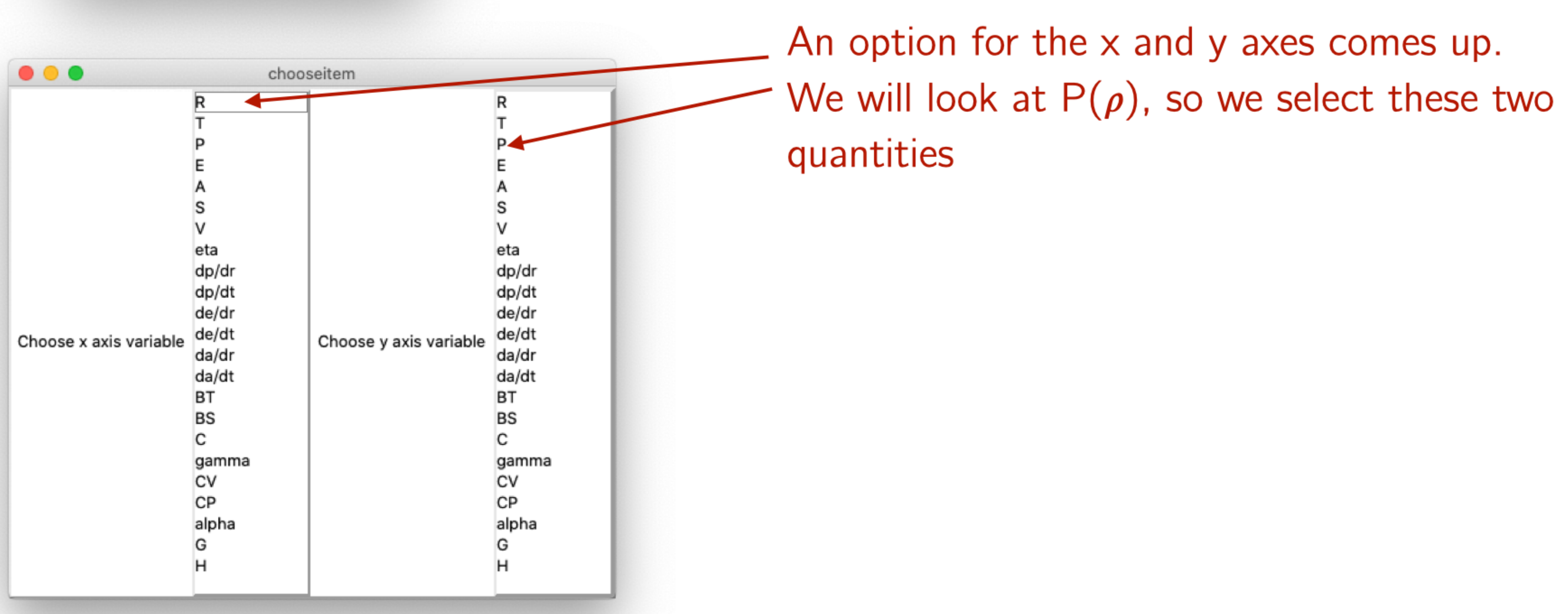} \end{center}
\begin{center}  \includegraphics[width=\textwidth]{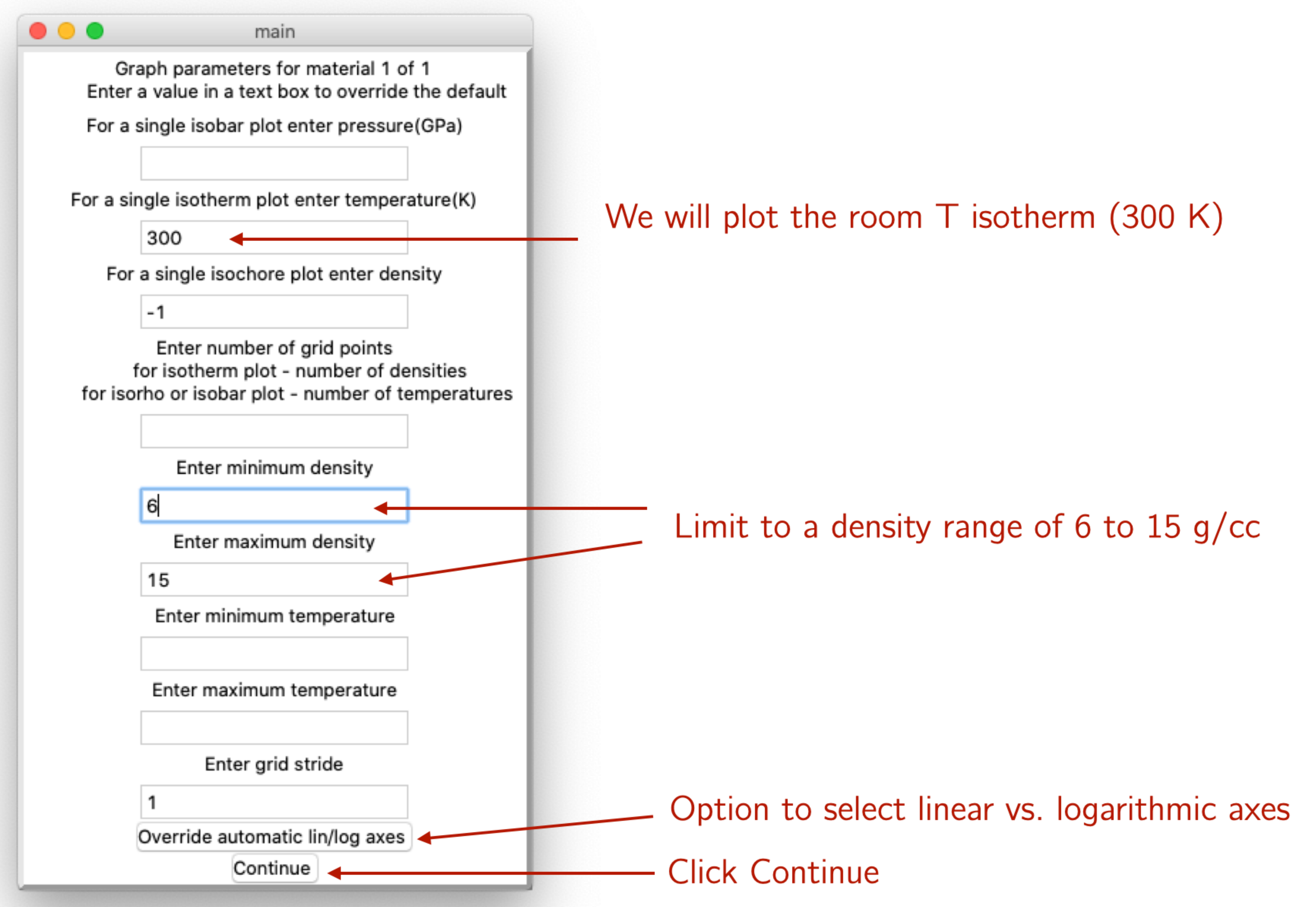} \end{center}
\begin{center}  \includegraphics[width=\textwidth]{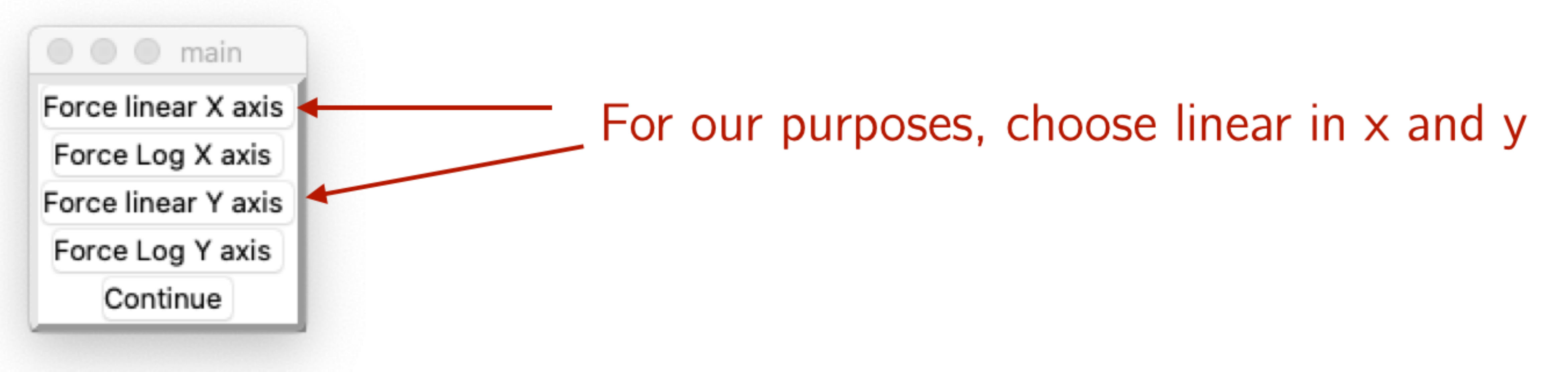} \end{center}
\begin{center}  \includegraphics[width=\textwidth]{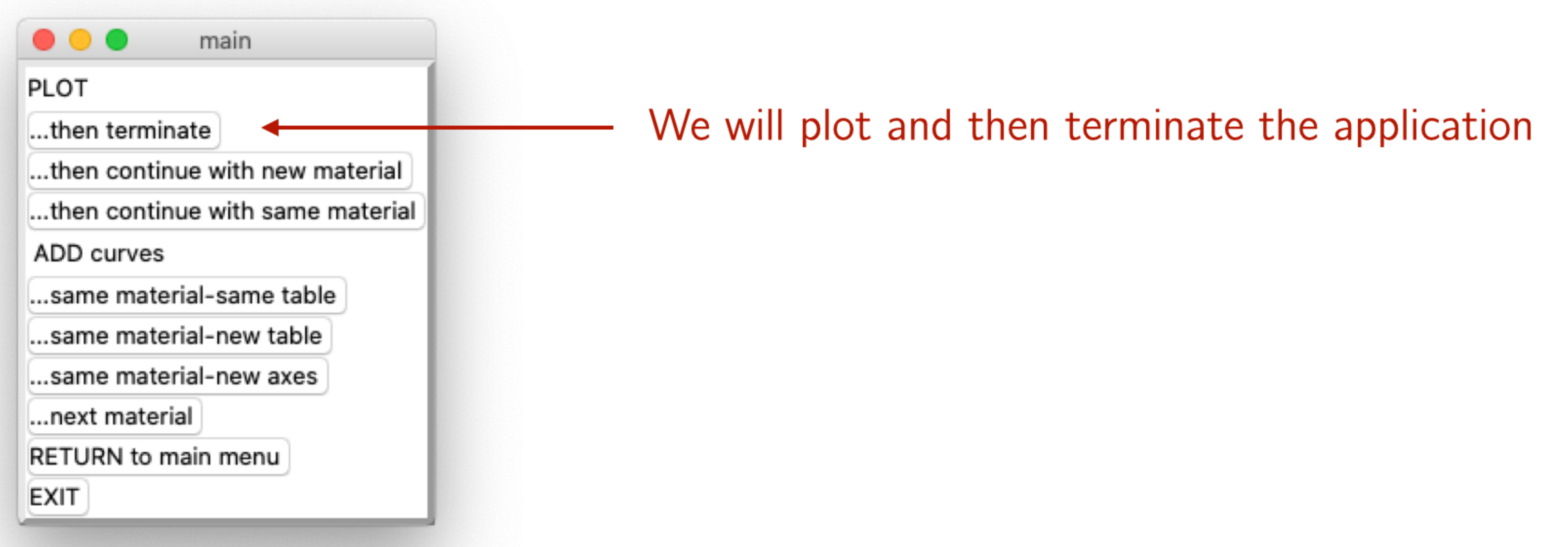} \end{center}
\begin{center}  \includegraphics[width=0.9\textwidth]{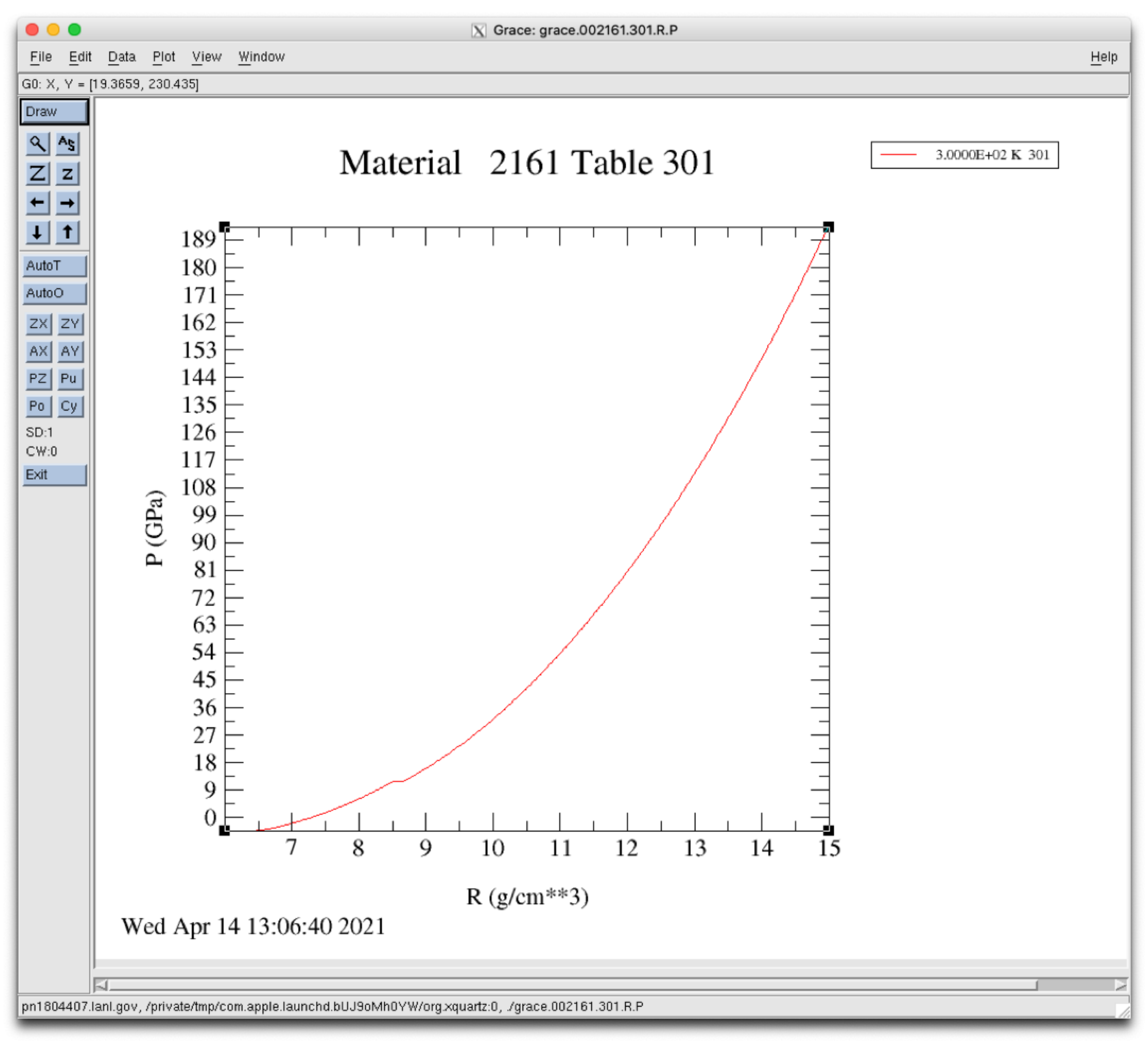} \end{center}

\newpage

After running the GUI, some files are automatically generated and
saved in the \texttt{gui} folder. Listing~\ref{lst:sesamein} shows the
contents of the \texttt{gui/sesamein.user} file created by the GUI.
The options we chose in the GUI are now stored in this file and we can
run opensesame on this file to create the isotherm plot. The
difference with this option from using the \texttt{xmgrace} file is
that if the contents of the 301 table in \texttt{lib/002161} have
changed, we will get a fresh plot of the new data. This is in contrast
to running \texttt{xmgrace} on the \texttt{grace.002161.301.R.P} file,
since that file contains the actual data points for the original
isotherm generated when running the GUI the first time. 
\begin{lstlisting}[language=Fortran,caption={\footnotesize\texttt{gui/sesamein.user}:
    auto-generated input file with selected GUI options},label={lst:sesamein}]
&job job_type='plot' sourcelib_path='../lib' /

&plot
 material_numbers=2161
 curve_type(1)='isotherm'
 curve_material(1)=1
 curve_table_number(1)=301
 curve_x_axis_name(1)='R'
 curve_y_axis_name(1)='P'
 curve_x_lin(1)='lin'
 curve_y_lin(1)='lin'
 curve_number_of_points(1)=0 
 curve_grid_lin(1)=.t.
 curve_t0(1)=300
 curve_rho0(1)=-1
 curve_rholo(1)=6
 curve_rhohi(1)=15 
 curve_nskip(1)=1 
 number_of_curves=1 
/
\end{lstlisting}

The benefit of the \texttt{sesamein.user} file is that we do not have
to rerun the GUI each time and select all of those options again, but
instead can run opensesame directly to get a fresh plot with the
updated EOS data:
\begin{Verbatim}
$ opensesame <  sesamein.user 
\end{Verbatim}

\newpage

If we prefer to use something other than \texttt{xmgrace} to plot
data, we can modify the sesamein.user file slightly to generate a data
file, shown in Listing~\ref{lst:sesamein2}.

\begin{lstlisting}[language=Fortran,caption={\footnotesize\texttt{gui/sesamein.user}:
    updated with options to save to a file},label={lst:sesamein2}]
&job job_type='plot' sourcelib_path='../lib' /

&plot
 output_device = 'data_file'     ! add save to file option
 plot_file_name = 'isotherm.dat' ! name of file to save to
 material_numbers=2161
 curve_type(1)='isotherm'
 curve_material(1)=1
 curve_table_number(1)=301
 curve_x_axis_name(1)='R'
 curve_y_axis_name(1)='P'
 curve_x_lin(1)='lin'
 curve_y_lin(1)='lin'
 curve_number_of_points(1)=0 
 curve_grid_lin(1)=.t.
 curve_t0(1)=300
 curve_rho0(1)=-1
 curve_rholo(1)=6
 curve_rhohi(1)=15 
 curve_nskip(1)=1 
 number_of_curves=1 
/
\end{lstlisting}

Here we have added these two lines with added comments that will make
save the isotherm to a file called \texttt{isotherm.dat}.

The file \texttt{isotherm.dat} has the form of Listing~\ref{lst:isotherm}.
\begin{lstlisting}[language=Python,caption={\footnotesize \texttt{gui/isotherm.dat}:
    isotherm data generated using save data option},label={lst:isotherm}]
# R "P  3.0000E+02 K  301"
  0.601201425000000D+01 -0.425908007663237D+01
  0.619419650000000D+01 -0.424562744668274D+01
  0.637637875000000D+01 -0.424452474215052D+01
  0.655856100000000D+01 -0.418344661032679D+01
  0.674074325000000D+01 -0.332385879461467D+01
  0.692292550000000D+01 -0.233533299185693D+01
  0.699579840000000D+01 -0.190578142291123D+01
  0.701562539427208D+01 -0.178551433934686D+01
  0.703545238854415D+01 -0.166376993441048D+01
  ...
\end{lstlisting}

\begin{figure}[!ht]\centering
  \includegraphics[width=0.7\textwidth]{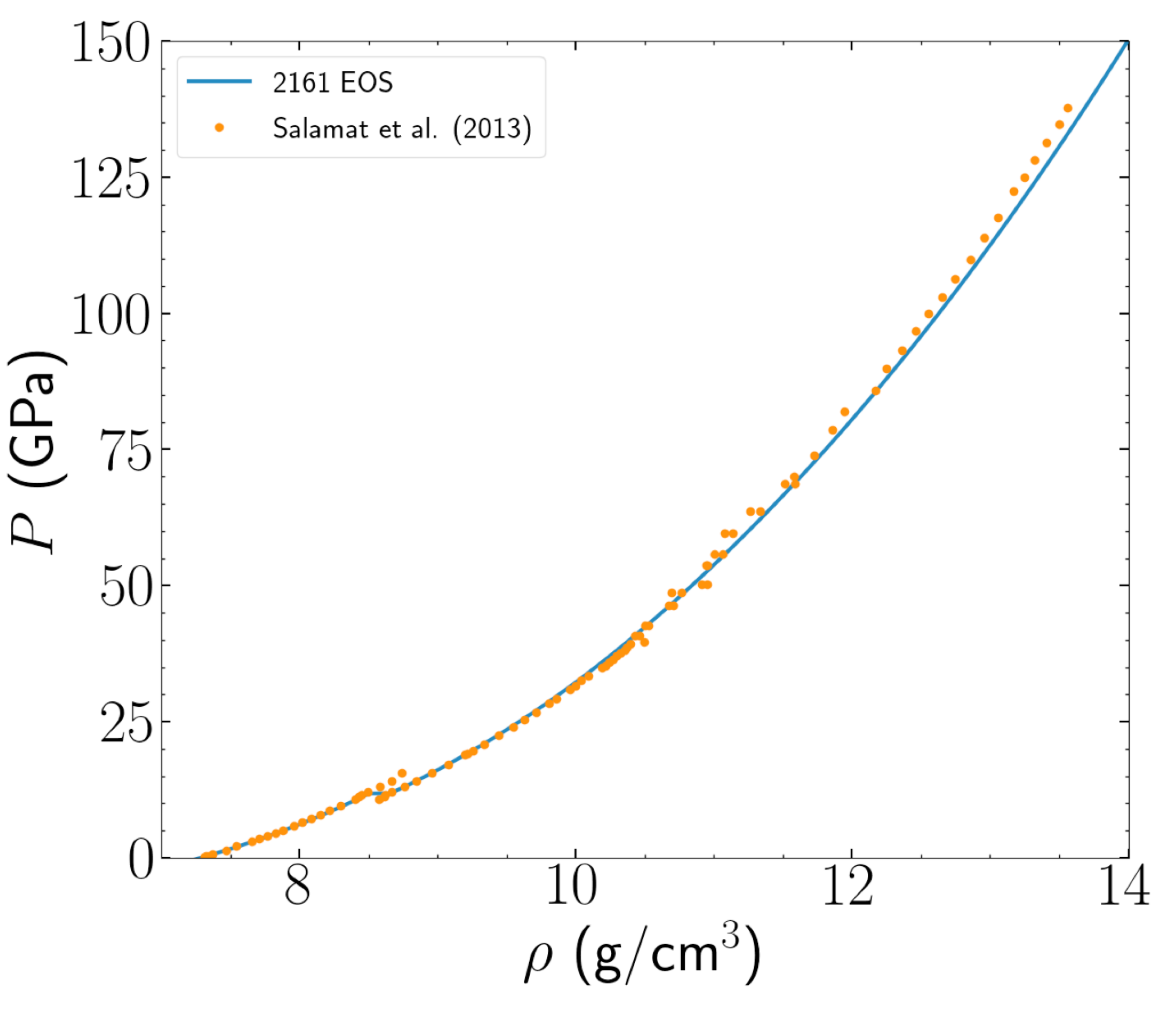}
  \caption{\footnotesize Isotherm of 2161 and data from Salamat et al. (2013)~\cite{salamat2013high}.}
  \label{fig:isotherm}
\end{figure}

Notice that the Fortran convention of using D for exponential format
is used.  This causes some problems if plotting the data in other
languages such as Python. A simple command to change the D to E using
\texttt{sed} is,
\begin{Verbatim}[fontsize=\footnotesize]
$  sed -i .bak 's/D/E/g'  isotherm.dat
\end{Verbatim}
With the \texttt{isotherm.dat} file, we can now use other languages
like Python to plot with other data. An example of a Python script
that plots the 2161 isotherm with a comparison to experimental data is
shown in Listing~\ref{lst:python}, which uses
matplotlib~\cite{matplotlib} and numpy~\cite{numpy}. The plot
generated from running this script is shown in
Fig.~\ref{fig:isotherm}.

\begin{lstlisting}[language=Python, caption={\footnotesize \texttt{gui/plot.dat}:
      Python script for plotting an isotherm with experimental data},
    label={lst:python}]
from pylab import *
import os
rcParams.update({'font.size':48, 'text.usetex': True})

os.system("sed -i .bak 's/D/E/g' isotherm.dat")

data = genfromtxt('isotherm.dat')      # load isotherm data
exp  = genfromtxt('2013-salamat.dat')  # load experimental data

figure(figsize=(14,12))
plot(data[:,0], data[:,1],lw=3, label='2161 EOS')
plot(exp[:,0], exp[:,1], 'o', label = 'Salamat et al. (2013)')
legend(fontsize=24)
xlim(7,14)
ylim(0,150)
ylabel('$P$ (GPa)')
xlabel(r'$\rho$ (g/cm$^3$)')
tight_layout()
show()
\end{lstlisting}

\newpage
\newpage

\section*{4. Creating ASCII-formatted EOS tables}
Typically, hydro code users use either ASCII-formatted EOS tables as
input to hydro codes.  To demonstrate how to do this, we create a new
folder titled \texttt{5-ascii}, so that we have

\begin{Verbatim}[fontsize=\footnotesize]
$ ls
1-beta/    2-gamma/    3-liquid/    4-multiphase/    5-ascii    lib/  
\end{Verbatim}

We then create another input file, \texttt{5-ascii/input.nml} with the
contents shown in Listing~\ref{lst:ascii}.  Running \texttt{opensesame
  < input.nml} in that folder will generate a file called
\texttt{ascii-2161} that can be shared with hydro code users.

\begin{lstlisting}[language=Fortran,caption={\footnotesize\texttt{5-ascii/input.nml}:
    file for generating ASCII formatted tables},label={lst:ascii}]
&job job_type = 'makeasciifile'
  sourcelib_path = '../lib/'  ! create ASCII using contents of lib folder
  resultlib_path = '../lib/'
  sesame_ascii_path = './ascii-2161' ! create ASCII file with name ascii-2161
/  
\end{lstlisting}

Note that with the ASCII file, it is possible to generate a
binary-formatted version of the ASCII file using EOSPAC.  We do not
cover that here since it involves the use of EOSPAC.

\newpage
\bibliographystyle{ieeetr}
\bibliography{refs}{}

\end{document}